\documentclass[11pt]{article}

\usepackage[final]{acl}

\usepackage{times}
\usepackage{latexsym}
\usepackage{booktabs}
\usepackage{multirow}
\usepackage{subcaption}
\usepackage{amsmath}
\usepackage{amssymb}
\usepackage{xcolor}
\usepackage[skins,breakable]{tcolorbox}
\newtcolorbox{simplebox}[1][]{
  breakable,
  colback=gray!3,
  colframe=black!70,
  boxrule=0.5pt,
  arc=2pt,
  left=8pt,
  right=8pt,
  top=8pt,
  bottom=8pt,
  fonttitle=\bfseries,
  title={#1}
}
\usepackage{enumitem}
\usepackage{ragged2e}
\usepackage[T1]{fontenc}

\usepackage[utf8]{inputenc}

\usepackage{microtype}

\usepackage{inconsolata}

\usepackage{graphicx}

\title{ATIR: Towards Audio-Text Interleaved Contextual Retrieval}

\author{
Tong Zhao, Chenghao Zhang, Yutao Zhu, Zhicheng Dou\thanks{Corresponding author.} \\
Gaoling School of Artificial Intelligence, Renmin University of China \\
\texttt{zhaotong7@ruc.edu.cn, dou@ruc.edu.cn} 
}

\begin{document}
\maketitle
\begin{abstract}
Audio carries richer information than text, including emotion, speaker traits, and environmental context, while also enabling lower-latency processing compared to speech-to-text pipelines. However, recent multimodal information retrieval research has predominantly focused on images, largely overlooking audio, especially in the setting of interleaved audio-text contextual retrieval. In this work, we introduce the Audio-Text Interleaved contextual Retrieval (ATIR) task, where queries can alternate between audio and text modalities. We construct an ATIR benchmark by integrating several Automatic Speech Recognition (ASR), QA, and retrieval datasets, ultimately unifying four types of contextual retrieval tasks. This benchmark substantially addresses the limitations of existing audio retrieval datasets in semantic retrieval. To study this task, we evaluate several off-the-shelf retrievers and train our ATIR model based on a Multimodal Large Language Model (MLLM). We further introduce a novel token compression mechanism that is orthogonal to existing compression methods, thereby alleviating the issue of excessive audio tokens in MLLM-based ATIR models. Experimental results demonstrate that our ATIR model achieves substantial improvements over strong baselines.
\end{abstract}

\section{Introduction}
\begin{figure}[t]
  \includegraphics[width=\columnwidth]{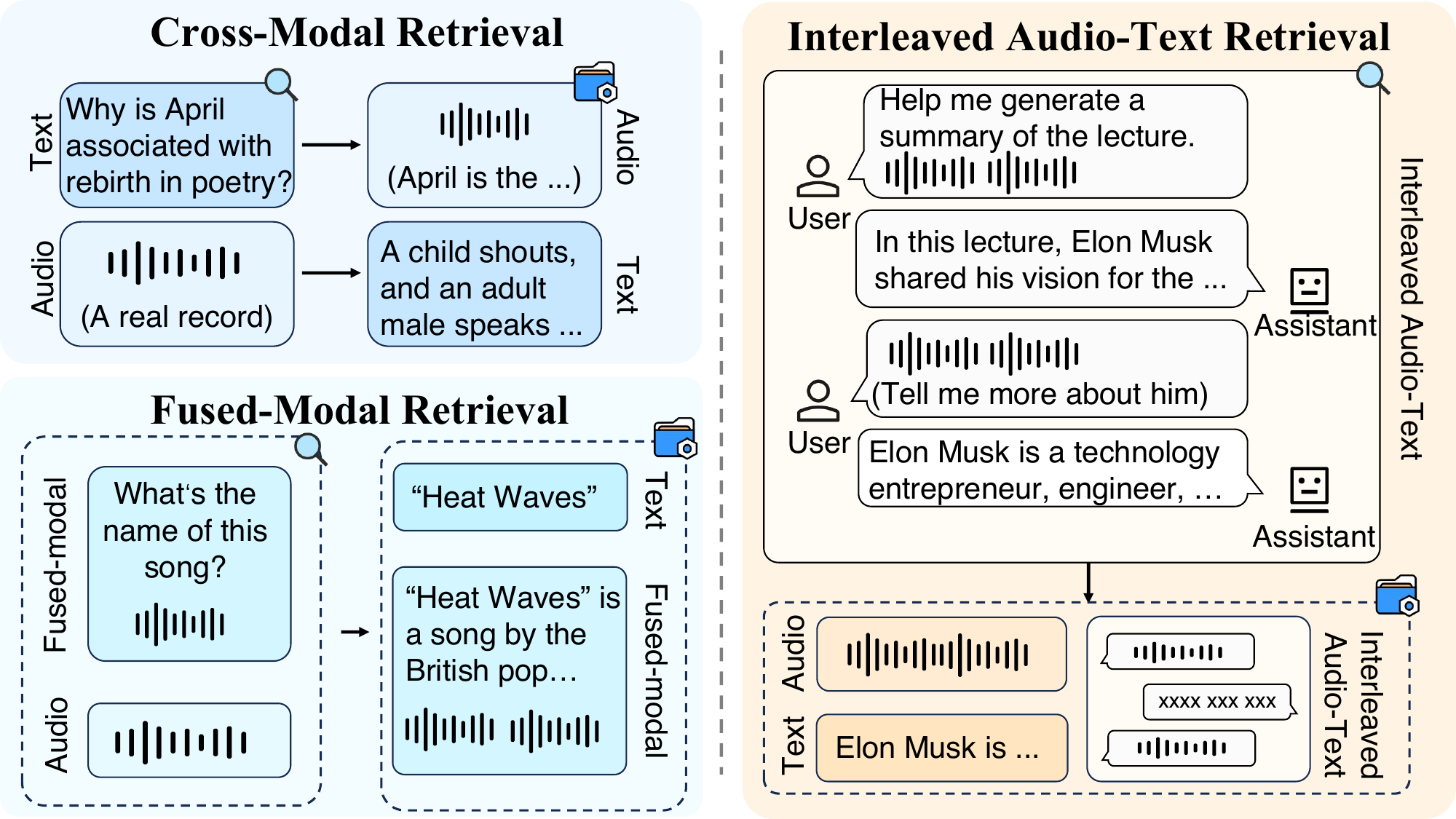}
  \caption{Comparison of traditional cross-modal and fused-modal retrieval settings with our proposed ATIR paradigm, where queries contain alternating segments that require contextual, multi-turn understanding.}
  \label{fig:intro}
\end{figure}

Multimodal information retrieval has emerged as a critical task that seeks to identify relevant information across heterogeneous data modalities~\citep{li2024improving,wei2024uniir,lan2025ume}. This field has advanced rapidly alongside the development of Multimodal Large Language Models (MLLMs)~\citep{xie2025show,deng2025emerging} and has found broad application in scenarios such as Retrieval-Augmented Generation (RAG) systems~\citep{chen2025wavrag,zhang2025mixedmodalretrievaluniversalretrievalaugmented}. Existing audio-text retrievers typically project audio and text into a shared embedding space, enabling both cross-modal retrieval and fused-modal retrieval tasks (Figure~\ref{fig:intro} left)~\citep{DBLP:conf/icassp/ElizaldeDW24,munakata2025language}, they primarily focus on static and single-turn interactions~\citep{munakata2025language,koepke2022audio}.

However, real-world communication is dynamic and inherently interleaved. In scenarios like conversational assistants~\citep{xu2025qwen2} and hybrid voice searches, users frequently switch between speaking and typing based on their environment. Similarly, content such as lecture recordings and meetings naturally combines coupled audio and textual information. These scenarios create a significant challenge: retrieval systems must process queries and documents where modalities are interleaved in a sequential and semantic order (as shown in the right side of Figure~\ref{fig:intro}). Existing retrievers fail to capture these complex contextual dependencies. Furthermore, directly applying MLLMs to this task is impractical. The significant difference in information density between audio and text leads to computational inefficiency, and the excessive length of audio tokens introduces noise that degrades retrieval accuracy.


To address these challenges, we introduce the Audio-Text Interleaved contextual Retrieval (ATIR) task. This task generalizes traditional retrieval paradigms by requiring models to understand the sequential and semantic relationships within alternating audio and text segments. To support this research, we construct a comprehensive benchmark derived from three diverse datasets: LibriSpeech~\citep{panayotov2015librispeech}, CoQA~\citep{reddy-etal-2019-coqa}, and SVQ~\citep{heigoldmassive}. We develop a rigorous synthesis pipeline to transform these sources into a unified ATIR format. This pipeline utilizes MLLMs to generate multi-turn, interleaved queries grounded in the source documents. To ensure the benchmark's difficulty and quality, we implement a strict self-evaluation mechanism and a hard negative mining strategy, filtering out low-quality samples to create a robust testbed for future research.


Beyond the benchmark, we propose a novel retrieval framework explicitly designed for the ATIR task. The core challenge in interleaved retrieval is balancing the rich semantics of text with the high redundancy of audio signals. Standard fine-tuning methods often allow audio tokens to dominate the embedding space, reducing performance. Our proposed method incorporates a token selector module within a bi-encoder architecture. This selector intelligently filters out redundant audio information, preserving only the most informative tokens. This approach balances information density across modalities, enabling efficient and accurate retrieval over long, audio–text interleaved sequences. Experimental results demonstrate that ATIR consistently outperforms strong baselines, with further ablation studies validating the effectiveness of each proposed component.

Our contributions can be summarized as follows:

(1) We formally define the ATIR task and identify the key limitations of existing single-modal and cross-modal retrievers in this context.

(2) We construct the first large-scale benchmark dedicated to audio-text interleaved retrieval, established through a rigorous data synthesis and quality control pipeline.

(3) We propose an ATIR-specific framework featuring a token selector that resolves the issues of audio redundancy and computational inefficiency. Extensive experiments show that our approach outperforms existing baselines and provides insights of modeling of complex multimodal sequences.

\section{Related Works}
\textbf{Multimodal Large Language Models.}
By expanding the capabilities of large language models beyond text, MLLMs enable the unified processing and reasoning of visual and audio signals alongside textual input, thereby supporting integrated multimodal understanding and generation.~\citep{chen2024mllm,DBLP:journals/corr/abs-2509-17765}.
BAGEL~\citep{DBLP:journals/corr/abs-2505-14683} shows that large-scale pretraining on interleaved multimodal data induces strong emergent multimodal reasoning in a unified decoder-only model. 
SALMONN~\citep{DBLP:conf/iclr/TangYSC000M024}, UniAudio~\citep{DBLP:conf/nips/YangGWHLTWM24}, and IntrinsicVoice~\citep{DBLP:journals/corr/abs-2410-08035} equip LLMs with speech and audio modeling capabilities, supporting tasks ranging from audio understanding to real-time speech interaction. 
VALOR~\citep{DBLP:journals/pami/LiuCHGZWT25} and WAVE~\citep{DBLP:journals/corr/abs-2509-21990} further explore unified representations across vision, audio, and language for multimodal understanding and retrieval, while Qwen3-Omni~\citep{xu2025qwen3} achieves outperforming performance across text, image, audio, and video within a single architecture. Despite these advances, most MLLMs focus on multimodal understanding or generation, with limited attention to retrieval-oriented representation learning for audio--text interleaved sequences. ATIR addresses this gap by learning unified embeddings for interleaved audio--text retrieval.

\noindent\textbf{Audio-Text Retrieval.} With the advancement of MLLMs, audio-text retrieval has gradually evolved from low-level feature matching to more contextual and semantically grounded retrieval. Early studies primarily focused on extracting discriminative audio representations and performing feature-based matching~\citep{mesaros2019sound,chen2022wavlm,baevski2020wav2vec}. CLAP~\citep{DBLP:conf/icassp/ElizaldeDW24} leverages contrastive language–audio pretraining to enable strong performance on a variety of audio classification tasks. \citet{gomes2022automated} generate audio captions and conduct retrieval over caption space, while \citet{oncescu2021audio,koepke2022audio} directly map audio and textual descriptions into a unified vector space for cross-modal retrieval. More recent work has shifted toward semantic and context-aware audio retrieval. \citet{chen2025wavrag} extend RAG frameworks to audio modalities, enabling generative systems to leverage audio evidence. \citet{munakata2025language} further explore language-driven audio moment retrieval. However, existing approaches still predominantly assume single-modality or single-turn inputs, lacking the ability to model complex interleaved audio–text structures that arise in real-world multimodal interactions.

\begin{figure*}[th]
  \includegraphics[width=\textwidth]{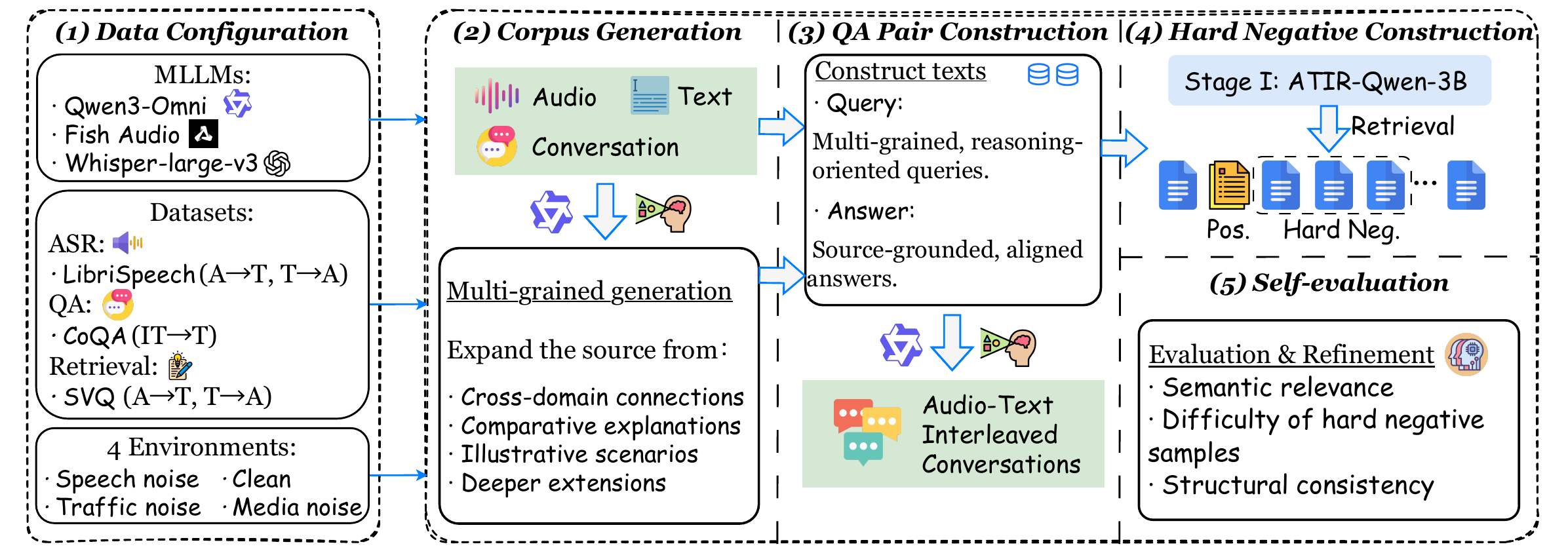}
  \caption{Overview of the ATIR dataset construction pipeline. The pipeline comprises data configuration and multi-grained corpus generation, query–answer pair construction with reasoning-oriented queries and source-grounded answers, hard negative mining for retrieval, and iterative self-evaluation for quality refinement.}
  \label{fig:dataset}
\end{figure*}

\section{ATIR Benchmark}
\subsection{Task Definition} \label{sec:task}
We define an \textit{audio--text interleaved} data instance $X$ as an ordered sequence of audio and text segments, denoted as $X = [x_1, \ldots, x_n],$ where each $x_i$ corresponds to either a text chunk or an audio segment, and the sequence is organized according to their contextual relationships. In ATIR settings, such an instance typically consists of multiple turns of audio--text interleaved dialogue, where audio and text segments alternate according to the flow of interaction. This form of interleaving arises naturally in conversations with MLLMs.

Given an interleaved query $X^{Q}$ and a corpus $\mathcal{C}=\{X^{D}_1, \ldots, X^{D}_m\}$, the ATIR task aims to retrieve the document $X^{D}$ in $\mathcal{C}$ that is most relevant to $X^{Q}$. Relevance is defined by a similarity function $s(X^{Q}, X^{D})$, which measures semantic similarity at the level of audio--text interleaved sequences. Different from conventional audio--text retrieval settings that typically assume single-turn or single-modality inputs, ATIR requires the model to understand contextually interleaved audio and text segments in both queries and documents, which could be challenging for existing multimodal retrievers.

\subsection{Data Synthesis Framework}
We construct a dedicated data synthesis framework to generate training data for adapting MLLMs to downstream embedding tasks. The design of this framework follows the core criteria for high-quality synthetic multimodal data established in prior work~\citep{DBLP:conf/acl/0005W00ZWD25}, with particular emphasis on broad coverage, reliable cross-modal correspondence, and high fidelity.

\subsubsection{Data Configuration}
As a first step, we configure the source data from three perspectives to support the subsequent synthesis procedure.

\paragraph{Task Types.} Our objective is to build synthesized data with broad task coverage, rather than limiting the construction process to conventional A$\rightarrow$T and T$\rightarrow$A retrieval settings. To this end, we draw the source data from three representative audio-related task families identified in previous studies~\citep{yu2016automatic,DBLP:journals/corr/abs-2304-13689,manning2008introduction}: ASR, QA, and Retrieval. For each selected dataset, we then apply the unified construction pipeline introduced in Section~\ref{sec:pipeline} to produce the final ATIR datasets.

\paragraph{Audio.} Existing datasets show inconsistent audio distributions across segments, for example, some datasets provide audio only for questions while others only for answers, which hinders the construction of unified audio--text interleaved structures. Meanwhile, recent MLLMs (\emph{e.g.}, GPT-4o and Qwen3-Omni~\citep{DBLP:journals/corr/abs-2509-17765}) are capable of generating high-fidelity and natural speech~\citep{yin2024survey}. We therefore employ MLLM to synthesize missing audio segments while preserving original recordings. This ensures structural consistency and reliable cross-modal alignment for ATIR.

\paragraph{Environments.} Most existing models are trained and evaluated primarily under a single acoustic condition, typically a clean setting without background noise. To synthesize data that reflect diverse real-world scenarios, we follow previous work~\citep{heigoldmassive} and synthesize audio under four representative acoustic environments: (1) \textit{clean}, where no additional noise is introduced; (2) \textit{background speech noise}, where speech signals from external sources such as podcasts or talk radio are mixed into the audio at a normal listening volume; (3) \textit{traffic noise}, where noise profiles corresponding to moving vehicles, including buses, trains, or cars, are added to simulate in-vehicle conditions; and (4) \textit{media noise}, where background media such as music, television, or movies is mixed into the audio at an audible yet natural level.

\subsubsection{Unified Synthesis Pipeline} \label{sec:pipeline}
With the data configuration in place, we introduce a unified synthesis pipeline that covers semantically relevant document generation, question--answer pair construction, hard negative construction, and self-evaluation. Although the sampled datasets originate from diverse formats, this pipeline transforms them into a unified dataset under the ATIR formulation that is introduced in Section~\ref{sec:task}.

\paragraph{Semantically Relevant Corpus Generation.} To obtain semantically rich and diverse related documents, the MLLM $M_\theta$ first expands the source content from multiple perspectives: (1) cross-domain connections to history, culture, or real-world applications, (2) comparative explanations or analogies with related concepts, (3) illustrative scenarios or concrete examples grounded in real-life contexts, and (4) deeper reasoning-oriented extensions, such as multi-hop inference, metaphorical interpretation, critical reflection, or temporal evolution. The multi-aspect expansion of the content enables the MLLM $M_\theta$ to generate a corpus that remains topically aligned with original text while exhibiting diverse discourse structures and reasoning patterns, thereby improving semantic coverage and supporting robust retrieval under the ATIR formulation.

\paragraph{Question--Answer Pair Construction.} To construct high-quality question--answer pairs, the MLLM $M_\theta$ generates questions that are grounded in the synthesized or origin corpus and require understanding of their core semantics. The corresponding answers are derived from the same documents, ensuring semantic consistency between the question and answer. This process produces aligned question--answer pairs that can be naturally integrated into audio--text interleaved sequences, supporting diverse retrieval scenarios in the ATIR benchmark.

\paragraph{Hard Negative Construction.} To improve the effectiveness of ATIR model training, we construct hard negative samples from two complementary aspects. (1) Following prior work on hard negative mining for retrieval~\citep{DBLP:journals/corr/abs-2504-17432}, we use the retriever trained in the first stage, which will be further introduced in Section~\ref{sec:multi-stage_trainging}, to retrieve candidate documents for each query. Retrieved samples ranked higher than the positive example are treated as false negatives, while those ranked below the positive but within the top retrieved results are selected as hard negatives. (2) We further adopt an MLLM-based generation approach to synthesize hard negatives, which consists of two steps: (a) conditioning the model on a query and its corresponding positive passage to maintain topical relevance, and (b) instructing the model to alter key facts, entities, or conclusions to generate fluent yet semantically misleading passages. By integrating hard negatives constructed from both aspects, we obtain a diverse and challenging negative set that effectively supports robust ATIR model training.

\paragraph{Self-evaluation.} To further improve the quality of the synthesized ATIR data, the MLLM $M_\theta$ performs self-evaluation from multiple aspects: (1) the semantic relevance between audio and text segments within each interleaved sequence, (2) the plausibility and difficulty of hard negative samples, (3) the structural consistency of the audio--text interleaving, and (4) the diversity of the synthesized content across different environments.

\begin{table}[t]
\centering
\small
\setlength{\tabcolsep}{3pt}
\renewcommand{\arraystretch}{1.1}
\begin{tabular}{lcccc}
\toprule
\multirow{2}{*}{\textbf{Part}}
& \multirow{2}{*}{\textbf{\#Samples}}
& \textbf{Avg.} & \textbf{Avg. Text} & \textbf{Avg.} \\
& & \textbf{\#Audio (s)} & \textbf{\#Tokens} & \textbf{\#Turns} \\
\midrule
Corpus   &  88283 & 101.30 & 262.31 & - \\
\midrule
Test Query  & 3909 & 5.17 & 21.58 & 2.02 \\
Train Query & 84374 & 7.28 & 29.22 & 1.96 \\
\bottomrule
\end{tabular}
\caption{Dataset statistics for our constructed ATIR dataset, including the number of samples and the average audio duration, text length, and interleaved turns. We count tokens by Qwen tokenizer.}
\label{tab:dataset_stats_single}
\end{table}

\subsection{Data Statistics}
The ATIR dataset consists of 88,283 annotated query-positive document pairs in total, as detailed in Table~\ref{tab:dataset_stats_single}. Of these, 3,909 pairs are reserved for evaluation, and the remaining 84,374 pairs are used for model training. Additional dataset analyses can be found in Appendix~\ref{appendix:data_stats}.

\begin{figure*}[th]
  \includegraphics[width=\textwidth]{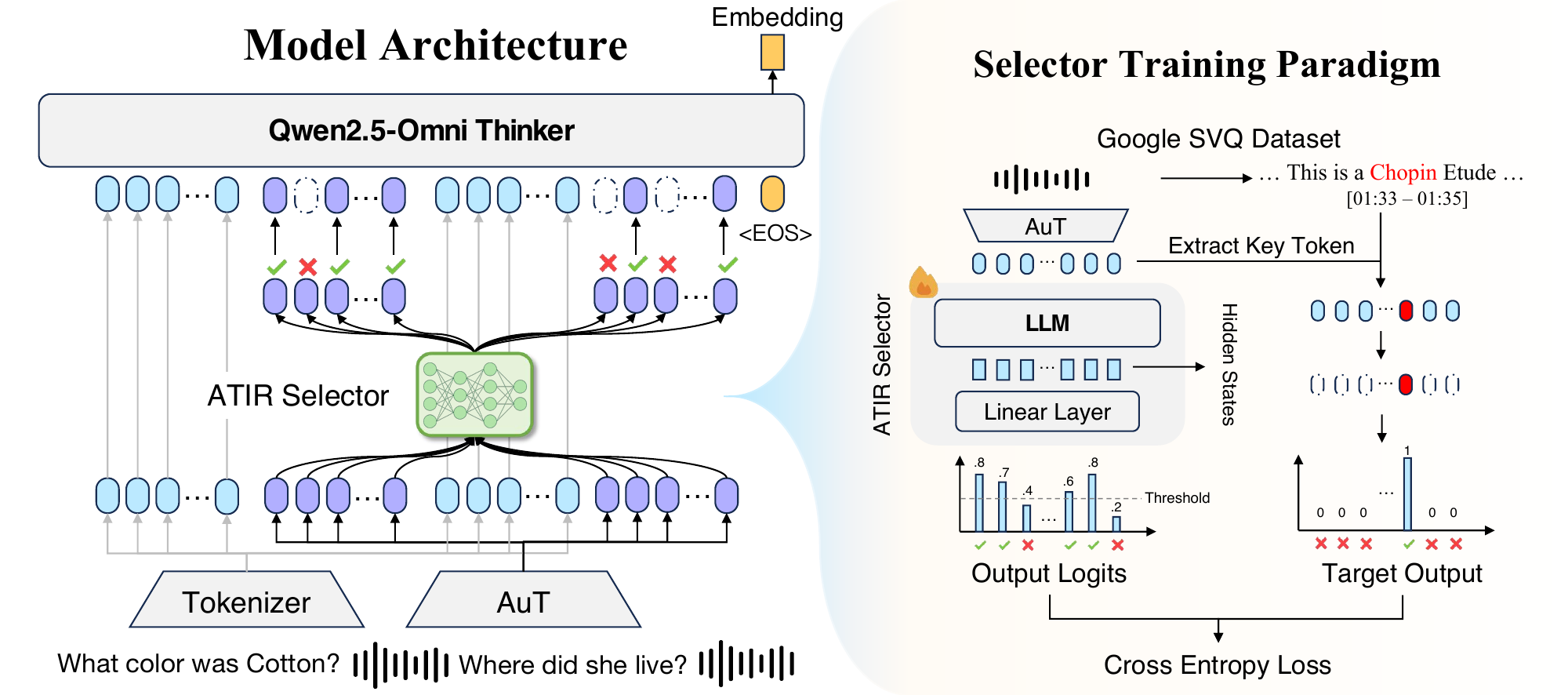}
  \caption{Architecture of the ATIR-Qwen-3B and the training paradigm of the ATIR Selector. The selector is plugged into the backbone to identify informative audio tokens and filter redundant context. It is trained with supervision derived from timestamp-level annotations, enabling selective context modeling and efficient audio--text interleaved representation learning.}
  \label{fig:embedding}
\end{figure*}
\section{Method}
\subsection{Model Architecture}
The ATIR model adopts a bi-encoder architecture, in which queries and documents are independently encoded into a shared embedding space, as illustrated in Figure~\ref{fig:embedding}. During inference, relevance is efficiently measured using similarity functions such as the dot product or cosine similarity. This architecture enables high scalability and supports efficient retrieval over large corpora.

An important feature of the ATIR model is its support for cross-modal, fused-modal, and interleaved-modal retrieval. Queries and corpus entries may originate from a single modality (e.g., text or audio) or from interleaved combinations of modalities (e.g., text+audio+text $\rightarrow$ audio+text+audio). This flexibility enables a wide range of retrieval scenarios, such as retrieving spoken explanations given mixed text--audio queries, matching multi-turn audio--text dialogues to relevant documents, and retrieving textual knowledge based on partially observed audio context.

We build our ATIR model on Qwen2.5-Omni-3B~\citep{DBLP:journals/corr/abs-2503-20215}, a foundation MLLM designed to process inputs across multiple modalities. Specifically, we adopt the Thinker backbone of the Qwen2.5-Omni architecture, which simplifies the overall design while retaining strong multimodal representation capabilities. We reuse the native audio encoder from Qwen2.5-Omni-3B, which follows the Qwen2-Audio encoder design. The audio encoder is kept fully frozen throughout both training stages, and only the remaining retrieval components are optimized. In our setup, each frame-level audio representation corresponds to approximately 40 ms of waveform. Since audio sequences can be long and often contain redundant or noisy segments, we introduce a selector module after the encoder to obtain more informative token sequences. 

\subsection{ATIR Selector} \label{sec:ATIR-selector}
Inspired by the selective context~\citep{DBLP:conf/emnlp/0001DGL23}, we propose a selector to obtain more informative token sequences, which is orthogonal to previous works that modify the tokenizer or encoder~\citep{DBLP:journals/corr/abs-2508-19205,DBLP:conf/iclr/Banerjee023}. The selector is built on Qwen3-0.6B~\citep{DBLP:journals/corr/abs-2505-09388}, a lightweight language model designed for efficient deployment while retaining strong understanding and generation capabilities. Concretely, the selector is implemented by adding a lightweight linear layer on top of the final hidden layer of the backbone model. As illustrated in Figure~\ref{fig:embedding}, this layer predicts a selection probability for each token position in the input sequence, indicating its importance for downstream retrieval. Tokens with probabilities above a predefined threshold are retained, while others are filtered out, resulting in a selective and compact context representation.

The proposed selector is a plug-and-play module that can be readily incorporated into different audio encoders and backbone models without any architectural changes. In addition, its adjustable threshold offers flexible control over the balance between contextual completeness and efficiency, enabling the selector to adapt to diverse ATIR scenarios with varying input lengths and noise conditions.

To supervise the selector, we leverage the SVQ dataset~\citep{heigoldmassive}, which provides audio recordings together with temporal annotations indicating the start and end timestamps of segments that contain salient information. Given an audio sequence and its corresponding timestamp annotations, we align the audio with the token sequence produced by the audio encoder and assign a binary supervision signal to each token.

Formally, let $\{h_1, \ldots, h_T\}$ denote the token representations output by the backbone model, where $T$ is the sequence length. For each token $h_t$, we define a binary label $y_t \in \{0,1\}$ indicating whether the token falls within an annotated informative time span. The selector predicts a selection probability $p_t \in [0,1]$ for each token via the linear layer. The selector is trained using a token-level binary classification objective:
\begin{equation}
\label{eq:selector_loss}
\mathcal{L}_{\text{sel}} = - \sum_{t=1}^{T} \left( y_t \log p_t + (1 - y_t) \log (1 - p_t) \right).\notag
\end{equation}

Through this supervision, the selector learns to identify and retain tokens corresponding to informative audio regions, while filtering out redundant or noisy context. Once trained, the selector can be applied to unseen data without timestamp annotations by thresholding the predicted probabilities.

\subsection{Model Training}
In this section, we present the key elements of the model training recipe, including the training objective and multi-stage training pipeline.

\subsubsection{Training Objective} We adopt the dense retrieval paradigm for training the ATIR model, where both interleaved queries and documents are encoded into a shared embedding space. The loss function that we utilized is the InfoNCE loss~\citep{oord2018representation}. Given an interleaved query $X^{Q}$, its corresponding relevant document $X^{D+}$, and a set of irrelevant documents $\{X^{D}_{i}\}_{i=1}^{N}$, the training objective is defined as:
\begin{equation}
\label{eq:infonce}
\mathcal{L} = - \log \frac{\exp\left(s(X^{Q}, X^{D+}) / \tau \right)}{\sum_{i=1}^{N} \exp\left(s(X^{Q}, X^{D}_{i}) / \tau \right)},
\end{equation}
where $s(\cdot,\cdot)$ denotes the cosine similarity between embeddings and $\tau$ is a temperature parameter. When hard negatives are not constructed, we use in-batch negatives.

This objective encourages the model to assign higher similarity scores to relevant audio--text interleaved pairs while effectively separating them from irrelevant candidates.

\subsubsection{Multi-stage Training} \label{sec:multi-stage_trainging}
Multi-stage training is a widely adopted strategy for training embedding models~\citep{DBLP:journals/corr/abs-2506-05176,DBLP:journals/corr/abs-2308-03281,DBLP:conf/acl/ChenXZLLL24,xu2025qwen3}. We follow this practice and design a two-stage training pipeline for ATIR, consisting of an embedding ability activation stage with weak supervision and an interleaved-modal capability elicitation stage. Both stages are trained using the same InfoNCE objective defined in Equation~\ref{eq:infonce}.

\paragraph{Stage I: Embedding Ability Activation}
In the first stage, we train the model using weakly supervised data constructed from single-modality and cross-modality pairs. This stage includes text--text, audio--audio, and audio--text pairs without explicit interleaving structures. The goal of this stage is to activate and stabilize the model’s representation learning ability across modalities, enabling it to learn coarse-grained semantic alignment before being exposed to more complex interleaved inputs.

\begin{table*}[t]
\centering
\scriptsize
\setlength{\tabcolsep}{4pt}
\renewcommand{\arraystretch}{1.1}
\resizebox{\textwidth}{!}{%
\begin{tabular}{c l | c c c c c c c c}
\toprule
\multirow{2}{*}{\textbf{Setting}}
& \multirow{2}{*}{\textbf{Model}}
& \multicolumn{2}{c}{\textbf{A$\rightarrow$T}}
& \multicolumn{2}{c}{\textbf{T$\rightarrow$A}}
& \multicolumn{2}{c}{\textbf{IAT$\rightarrow$T}}
& \multicolumn{2}{c}{\textbf{IAT$\rightarrow$A}} \\
\cmidrule(lr){3-4}\cmidrule(lr){5-6}\cmidrule(lr){7-8}\cmidrule(lr){9-10}
 &  & \textbf{R@1} & \textbf{nDCG@5}
     & \textbf{R@1} & \textbf{nDCG@5}
     & \textbf{R@1} & \textbf{nDCG@5}
     & \textbf{R@1} & \textbf{nDCG@5} \\
\midrule

\multirow{3}{*}{\textbf{Text}}
 & BGE-large
 & 65.33 & 74.16
 & 62.26 & 70.68
 & 61.36 & 69.57
 & 55.81 & 63.45 \\
  & E5-large
 & 66.20 & 74.83
 & 66.64 & 75.23
 & 66.44 & 73.04
 & 58.88 & 65.30 \\
 & Qwen3-Embedding-4B
 & \underline{76.64} & \underline{84.03}
 & 67.81 & 77.24
 & 69.24 & 76.37
 & 63.65 & 73.10 \\
\midrule

\multirow{3}{*}{\textbf{Cross}}
 & CLAP
 & 20.87 & 31.64
 & 13.46 & 19.63
 & 12.45 & 19.63
 & 11.13 & 17.79 \\
 & LAION-CLAP
 & 16.04 & 26.07
 & 11.25 & 17.48
 & 9.81 & 14.52
 & 8.89 &  12.27\\
 & M2D-CLAP
 & 34.42 & 49.79
 & 23.77 & 32.86
 & 22.53 & 30.58
 & 20.54 & 27.69 \\
\midrule

\multirow{2}{*}{\textbf{Fused}}
 & Omni-Embed-Nemotron-3B\footnotemark[2]
 & 73.28 & 81.56
 & \underline{72.54} & \underline{80.23}
 & 75.47 & 81.02
 & 64.49 & 77.63 \\
 & ColQwen-Omni-3B
 & 69.85 & 78.89
 & 71.61 & 79.33
 & \underline{79.69} & \underline{85.46}
 & \underline{68.64} & \underline{80.79} \\
 \midrule
 \textbf{Interleaved}
 & ATIR-Qwen-3B
 & \textbf{84.69} & \textbf{89.27}
 & \textbf{74.67} & \textbf{80.59}
 & \textbf{81.74} & \textbf{87.88}
 & \textbf{74.34} & \textbf{82.61} \\
\bottomrule
\end{tabular}}
\caption{Evaluation results on the ATIR benchmark, where \textbf{R@1} denotes Recall@1. We report Recall@1 and nDCG@5 for all retrieval settings. The best results are shown in \textbf{bold}, and the second-best are \underline{underlined}.}
\label{tab:ATIR_results}
\end{table*}

\paragraph{Stage II: Interleaved-modal Capability Elicitation}
In the second stage, we further train the model on audio--text interleaved data that explicitly contains strong and hard negative samples. The queries and documents in this stage involve alternating sequences of audio and text segments, closely matching the ATIR setting. By introducing interleaved-modal structures and challenging negatives, this stage strengthens the model’s ability to perform fine-grained interleaved-modal retrieval and improves robustness in complex retrieval scenarios.

\section{Experiments}
\subsection{Evaluated Models}
We adapt several kinds of retrievers for evaluation:
\begin{itemize}
    \item \textbf{Text models}, \emph{i.e.}, E5~\citep{DBLP:journals/corr/abs-2402-05672}, BGE~\citep{xiao2024c}, and Qwen3-Embedding-4B~\citep{DBLP:journals/corr/abs-2506-05176}. To evaluate these models, we replace audio segments with text transcriptions produced by the ASR model Whisper-Large-V3~\citep{DBLP:conf/icml/RadfordKXBMS23}.
    \item \textbf{Cross-modal models}, \emph{i.e.}, CLAP~\citep{DBLP:conf/icassp/ElizaldeDW24}, LAION-CLAP~\citep{DBLP:conf/icassp/ElizaldeDIW23}, and M2D-CLAP~\citep{DBLP:journals/access/NiizumiTYNOH25}. We evaluate these models under the same ASR-based protocol by converting interleaved audio segments into text using Whisper-Large-V3.
    \item \textbf{Fused-modal models}, \emph{i.e.}, ColQwen-Omni-3B~\citep{DBLP:conf/iclr/FaysseSWOVHC25} and the fine-tuned Omni-Embed-Nemotron-3B~\citep{DBLP:journals/corr/abs-2510-03458}. For these models, we concatenate interleaved text segments into a single text sequence and merge audio segments into a single audio input to form fused-modal representations.
\end{itemize}

\subsection{Settings}
\paragraph{Metrics.} We evaluate test-set performance using Recall@k, which measures the proportion of queries whose positive document is retrieved within the top-$k$ results, and nDCG@k (normalized Discounted Cumulative Gain), which assesses ranking quality by jointly considering the relevance and rank positions of positive results within the top-$k$.


\subsection{Main Results}
\newcommand{\yes}{\textcolor{green!60!black}{\ensuremath{\checkmark}}}
\newcommand{\no}{\textcolor{red!70!black}{\ensuremath{\times}}}
\begin{table*}[t]
\centering
\small
\scriptsize
\setlength{\tabcolsep}{3pt}
\renewcommand{\arraystretch}{1.1}
\resizebox{\textwidth}{!}{
\begin{tabular}{ccc c c c c c c c c c c}
\toprule
\multirow{2}{*}{\textbf{Sel.}}
& \multirow{2}{*}{\textbf{Stage I}}
& \multirow{2}{*}{\textbf{Stage II}}
& \multicolumn{2}{c}{\textbf{A$\rightarrow$T}}
& \multicolumn{2}{c}{\textbf{T$\rightarrow$A}}
& \multicolumn{2}{c}{\textbf{IAT$\rightarrow$T}}
& \multicolumn{2}{c}{\textbf{IAT$\rightarrow$A}}
& \multirow{2}{*}{$\boldsymbol{\Delta}$\textbf{R@1}}
& \multirow{2}{*}{$\boldsymbol{\Delta}$\textbf{nDCG@5}} \\
\cmidrule(lr){4-5}\cmidrule(lr){6-7}\cmidrule(lr){8-9}\cmidrule(lr){10-11}
 &  &  & \textbf{R@1} & \textbf{nDCG@5}
 & \textbf{R@1} & \textbf{nDCG@5}
 & \textbf{R@1} & \textbf{nDCG@5}
 & \textbf{R@1} & \textbf{nDCG@5}
 &  & \\
\midrule
\yes & \yes & \yes
& 84.69 & 89.27
& 74.67 & 80.59
& 81.74 & 87.88
& 74.34 & 82.61
& 0.00  & 0.00 \\

\no & \yes & \yes
& 83.89 & 88.12
& 73.96 & 79.27
& 80.66 & 86.31
& 72.92 & 80.97
& -1.05 & -1.42 \\

\no & \no & \yes
& 82.15 & 86.89
& 72.34 & 77.85
& 79.21 & 84.76
& 71.58 & 79.43
& -3.27 & -3.92 \\

\no & \yes & \no
& 80.42 & 85.23
& 70.89 & 76.14
& 77.65 & 82.94
& 69.83 & 77.65
& -5.86 & -6.75 \\
\bottomrule
\end{tabular}}
\caption{Ablation study of ATIR-Qwen-3B under different training configurations. \textbf{Sel.} indicates whether the ATIR Selector is applied, while \textbf{Stage I} and \textbf{Stage II} denote the two-stage training pipeline. Results are reported for four retrieval settings. $\Delta$R@1 and $\Delta$nDCG@5 represent the average performance change relative to the full model across all retrieval settings.}
\label{tab:ablation_router_r3_component}
\end{table*}
\footnotetext[2]{We finetuned Omni-Embed-Nemotron-3B, which will be further expounded in Appendix~\ref{appendix:implementation_details}.}

\paragraph{Comparison with Text-modal Models.} Table~\ref{tab:ATIR_results} reports the test results for a range of text-modal embedding models. For a fair comparison, we simplify the task for these models by replacing audio segments with transcriptions generated by Whisper-Large-V3. Despite this favorable setting, ATIR-Qwen-3B consistently achieves the best performance, with an average Recall@1 of 78.86\% and an average nDCG@5 of 85.09\%. It outperforms the strongest text-modal baseline, Qwen3-Embedding-4B (69.34\% Recall@1 and 77.69\% nDCG@5), by +9.52\% and +7.40\%, respectively. These results demonstrate that directly modeling audio--text interleaved inputs is more effective than the traditional ASR-then-embedding pipeline.
\paragraph{Comparison with Cross-modal Models.}
As shown in Table~\ref{tab:ATIR_results}, cross-modal models perform poorly on ATIR across all settings. The strongest baseline, M2D-CLAP, achieves only 34.42\% Recall@1 and 49.79\% nDCG@5 on A$\rightarrow$T, and further drops to 22.53\% / 30.58\% on IAT$\rightarrow$T. This is expected, as these models are mainly trained for audio--caption retrieval with short descriptive text, which does not transfer well to semantic retrieval with interleaved context.

\paragraph{Comparison with Fused-modal Models.}
ATIR-Qwen-3B consistently outperforms the best fused-modal baselines across all settings. It surpasses Omni-Embed-Nemotron-3B on A$\rightarrow$T by +11.41\% Recall@1 and +7.71\% nDCG@5, and improves over the strongest baselines on interleaved retrieval, e.g., +2.05\% / +2.42\% on IAT$\rightarrow$T and +5.70\% / +1.82\% on IAT$\rightarrow$A. These results indicate that collapsing interleaved inputs into a single sequence limits fused-modal models, while ATIR benefits from explicitly modeling audio--text interleaving.

\subsection{Ablation and Efficiency Analysis} \label{sec:ablation_and_efficiency}
\begin{table}[t]
\centering
\small
\setlength{\tabcolsep}{2pt}
\renewcommand{\arraystretch}{1.15}
\begin{tabular}{lcc}
\toprule
\textbf{Retriever} & \textbf{Params (B)} & \textbf{Latency (ms)} \\
\midrule
ASR + BGE-large           & 0.335 & 526.3 \\
ASR + E5-large            & 0.560 & 527.1 \\
ASR + Qwen3-Embedding-4B  & 4.021 & 531.9 \\
ASR + CLAP                      & 0.196 & 263.1 \\
ASR + LAION-CLAP                & 0.154 & 287.5 \\
ASR + M2D-CLAP                  & 0.089 & 265.7 \\
Omni-Embed-Nemotron-3B    & 4.703 & 19.5 \\
ColQwen-Omni-3B           & 4.396 & 17.1 \\
ATIR-Qwen-3B              & 4.396 & 16.8 \\
\bottomrule
\end{tabular}
\caption{Model size and inference latency of different retrievers. Latency is measured as the average end-to-end embedding time per query under the four types of evaluation setup.}
\label{tab:retriever_params_latency}
\end{table}
\begin{figure}[t]
  \includegraphics[width=\columnwidth]{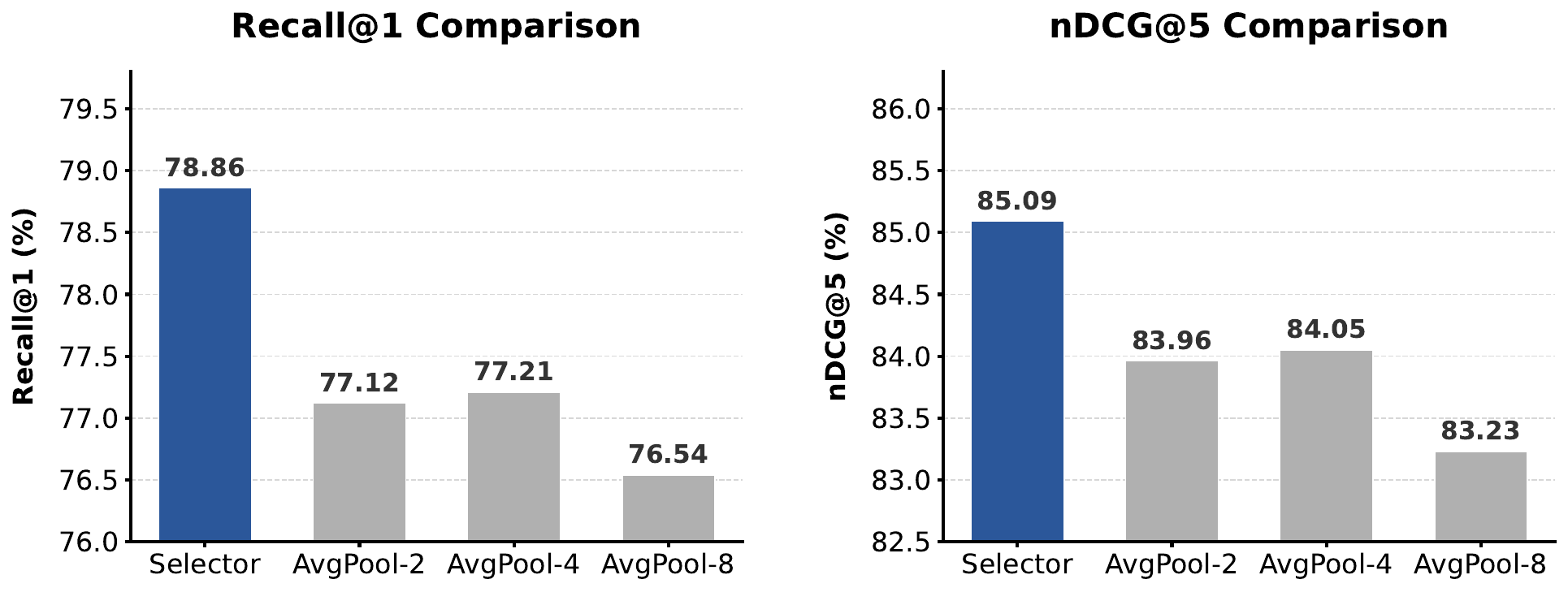}
  \caption{Comparison between the ATIR Selector and average pooling over audio tokens, reported by average Recall@1 and nDCG@5 across all retrieval settings.}
  \label{fig:comparison}
\end{figure}

\begin{table}[t]
\centering
\small
\begin{tabular}{ccccc}
\toprule
WER$\downarrow$ & CER$\downarrow$ & Sent.\ Acc$\uparrow$ & Word Acc$\uparrow$ & Char Acc$\uparrow$ \\
\midrule
0.0281 & 0.0093 & 0.6962 & 0.9719 & 0.9907 \\
\bottomrule
\end{tabular}
\caption{Normalized ASR quality of Whisper-Large-V3 on the ATIR test set.}
\label{tab:asr_quality}
\end{table}

\begin{table*}[t]
\centering
\small
\setlength{\tabcolsep}{4.5pt}
\begin{tabular}{llcccccccc}
\toprule
\multirow{2}{*}{Method} & \multirow{2}{*}{Input} &
\multicolumn{2}{c}{A$\rightarrow$T} &
\multicolumn{2}{c}{T$\rightarrow$A} &
\multicolumn{2}{c}{IAT$\rightarrow$T} &
\multicolumn{2}{c}{IAT$\rightarrow$A} \\
\cmidrule(lr){3-4} \cmidrule(lr){5-6} \cmidrule(lr){7-8} \cmidrule(lr){9-10}
& & R@1 & nDCG@5 & R@1 & nDCG@5 & R@1 & nDCG@5 & R@1 & nDCG@5 \\
\midrule
BGE-large           & ASR & 65.33 & 74.16 & 62.26 & 70.68 & 61.36 & 69.57 & 55.81 & 63.45 \\
BGE-large           & Src & 66.27 & 75.05 & 66.42 & 74.89 & 63.58 & 71.89 & 63.64 & 71.92 \\
E5-large            & ASR & 66.20 & 74.83 & 66.64 & 75.23 & 66.44 & 73.04 & 58.88 & 65.30 \\
E5-large            & Src & 68.93 & 79.12 & 68.82 & 79.04 & 69.15 & 77.34 & 69.22 & 77.45 \\
Qwen3-Embedding-4B  & ASR & 76.64 & 84.03 & 67.81 & 77.24 & 69.24 & 76.37 & 63.65 & 73.10 \\
Qwen3-Embedding-4B  & Src & 77.79 & 85.13 & 77.53 & 84.89 & 70.47 & 78.21 & 70.32 & 78.33 \\
\bottomrule
\end{tabular}
\caption{Controlled comparison of text retrievers using ASR transcripts versus oracle source text (``Src''). }
\label{tab:asr_oracle}
\end{table*}

\paragraph{Effects of ATIR-Qwen-3B Training Configurations.} As shown in Table~\ref{tab:ablation_router_r3_component}, removing the ATIR Selector results in a consistent performance drop (-1.05\% Recall@1 and -1.42\% nDCG@5 on average), demonstrating the benefit of selective audio token filtering. Disabling Stage~I leads to a larger degradation (-3.27\% / -3.92\%), while removing Stage~II causes the most severe drop (-5.86\% / -6.75\%), highlighting the importance of interleaved-modal training with strong negatives. Overall, all components contribute complementarily to ATIR-Qwen-3B.

\paragraph{Efficiency Analysis.}
As shown in Table~\ref{tab:retriever_params_latency}, the reported latency of text-based retrievers includes ASR overhead, as audio inputs must be transcribed for all retrieval settings; cross-modal retrievers similarly incur ASR costs for interleaved retrieval. In contrast, fused-modal and interleaved-modal retrievers process audio directly, avoiding ASR latency. With a comparable model size, ATIR-Qwen-3B achieves the lowest latency, demonstrating that selective audio modeling enables both efficient and effective audio--text interleaved retrieval.

\paragraph{ATIR Selector vs. Average Pooling.}
We compare the ATIR Selector with standard average pooling over audio tokens using $k\!\in\!\{2,4,8\}$ segments. ATIR-Qwen-3B achieves 78.86\% Recall@1 and 85.09 nDCG@5, outperforming 2/4/8-way average pooling (77.12/77.21/76.54 Recall@1 and 83.96/84.05/83.23 nDCG@5). These results show that uniform averaging is sensitive to noise, while the ATIR Selector better preserves informative audio content for interleaved retrieval.

\begin{table*}[t]
\centering
\small
\begin{tabular}{lcccc}
\toprule
Setting & IAT$\rightarrow$T R@1 & IAT$\rightarrow$T nDCG@5 & IAT$\rightarrow$A R@1 & IAT$\rightarrow$A nDCG@5 \\
\midrule
Original         & 81.74 & 87.88 & 74.34 & 82.61 \\
Shuffle Order    & 80.98 & 87.21 & 73.61 & 81.94 \\
Shuffle Position & 80.29 & 86.53 & 72.95 & 81.26 \\
Shuffle Both     & 79.54 & 85.79 & 72.23 & 80.51 \\
\bottomrule
\end{tabular}
\caption{Impact of disrupting interleaving structure by shuffling audio order and/or audio positions. }
\label{tab:interleaving_perturbation}
\end{table*}

\paragraph{Disentangling ASR Errors from Retrieval Quality}
Since text-only retrievers cannot directly process audio, we evaluate them using Whisper-Large-V3 transcripts. While practical, this protocol may conflate retrieval quality with transcription errors. To disentangle the two, we conduct two analyses. First, under a per-segment evaluation aligned with the interleaved turn structure, Whisper-Large-V3 achieves a normalized WER of 2.81\% on our test set (Table~\ref{tab:asr_quality}), indicating strong transcription quality. Second, we compare the same text retrievers using ASR transcripts versus oracle source text. As shown in Table~\ref{tab:asr_oracle}, oracle text consistently improves all text baselines, especially on T$\rightarrow$A and interleaved retrieval, while preserving their relative ranking. This suggests that transcription noise is non-negligible, but does not fully account for the gap between ASR-based pipelines and ATIR-Qwen-3B.

\paragraph{Impact of Interleaving Structure.}
To verify that ATIR depends on coherent interleaving structure rather than merely the presence of both modalities, we perform a controlled perturbation study that keeps the content unchanged while disrupting the alignment between audio and text turns. We consider three perturbations: \textit{Shuffle Order}, \textit{Shuffle Position}, and \textit{Shuffle Both}. As shown in Table~\ref{tab:interleaving_perturbation}, all perturbations degrade retrieval performance, with larger drops under stronger perturbations. This suggests that ATIR depends on structured sequential alignment between alternating modalities, rather than simple audio-text fusion.

\section{Conclusion}
We introduce audio-text interleaved contextual retrieval (ATIR) and construct the first benchmark for this task through an automatic synthesis pipeline over multiple audio and text datasets. To address the challenges of modeling long or noisy interleaved inputs, we propose ATIR-Qwen-3B, an interleaved-modal retriever with selective audio token filtering. Extensive experiments show that ATIR-Qwen-3B consistently outperforms text-modal, cross-modal, and fused-modal baselines, while achieving lower inference latency by avoiding ASR pipeline. Ablation and efficiency analyses further validate the effectiveness of selective audio modeling and multi-stage training. We hope this work will stimulate future research on retrieval-oriented multimodal representation learning.

\section{Limitations}
Despite demonstrating outperforming performance on the ATIR benchmark and validating the effectiveness of audio--text interleaved retrieval, several aspects remain open for future exploration. First, motivated by efficiency and scalability considerations, the current framework adopts a relatively lightweight representation design; exploring more expressive modeling strategies may further improve retrieval performance while balancing computational cost. Second, although the model handles interleaved queries effectively, it focuses on retrieving a single relevant document and does not consider more complex retrieval settings that combine evidence from multiple contexts. Finally, the evaluation is conducted on QA-centric ATIR tasks, and extending the framework to a broader range of multimodal tasks and application scenarios would help better assess its generality.


\bibliography{custom}

\appendix

\section{ATIR Benchmark Details}
\label{sec:details_of_ATIR_benchmark}
\subsection{Mapping Source Tasks to ATIR Format}

We design task-specific mapping strategies to transform heterogeneous source datasets into a unified ATIR format. Although ASR, QA, and retrieval datasets differ in their original structures, all of them are ultimately processed through the unified synthesis pipeline described in Section~\ref{sec:pipeline}, with differences mainly in the construction of initial interleaved sequences.

\paragraph{ASR Dataset.}
For LibriSpeech, the original data consist only of speech recordings and their corresponding transcriptions. We therefore apply the full synthesis pipeline, including corpus generation, question--answer pair construction, hard negative construction, and self-evaluation. The original transcriptions are used as source content to generate semantically related documents, enabling the transformation of single-turn ASR data into ATIR-style multi-turn audio--text interleaved instances.

\paragraph{QA Dataset.}
For CoQA, which already contain multi-turn question--answer dialogues grounded in a shared story, we reuse the original dialogue turns as the backbone of the interleaved sequence. The story and dialogue context are used to generate the corpus, after which the remaining stages of the pipeline are applied without modification.

\paragraph{Retrieval Dataset.}
For SVQ, which are typically single-turn in nature, we synthesize multi-turn dialogues to better reflect realistic retrieval interactions. Starting from the original query--document pairs, additional conversational turns are generated and interleaved with audio and text, followed by the standard pipeline procedures.

\subsection{Details of Corpus Generation Prompt Design}

To construct a semantically rich corpus, we employ prompt-based content expansion that guides the MLLM to generate related documents from multiple perspectives described in the main paper. All prompts explicitly discourage direct rephrasing and require the generated text to remain concise and factually relevant. Representative prompts used in corpus generation are shown below.

\begin{tcolorbox}[
  breakable,
  enhanced,
  colback=gray!2,           
  colframe=black!70,        
  boxrule=0.5pt,            
  arc=2pt,                  
  left=10pt, right=10pt, top=12pt, bottom=10pt,
  attach boxed title to top left={yshift=-3mm, xshift=5mm},
  boxed title style={
    sharp corners, 
    size=small, 
    colback=black!70, 
    colframe=black!70
  },
  title=\textbf{Representative Corpus Generation Prompts},
  coltitle=white,
  fonttitle=\footnotesize\sffamily,
  fontupper=\small\ttfamily\linespread{1.1}\selectfont, 
]

You are an expert writer who connects concepts across different fields.
Based on the following content, generate a related document that makes
connections to history, culture, or real-world applications.
Avoid direct rephrasing.

Original content: \{content\}

The document should be natural and under 150 words.

You are a knowledgeable author skilled in comparative writing.
Based on the following content, write a document that explains the
topic by comparing it with a different but related concept, or by using
an analogy. Avoid simple restatement.

Original content: \{content\}

Keep it under 150 words.

You are an expert in reasoning. Based on the following content, create a
document that not only explains the topic but also connects it with a
second-level related idea, requiring at least one reasoning step.
Do not simply paraphrase.

Original content: \{content\}

Write under 150 words.

You are a historian and futurist. Based on the following content, generate
a document that either imagines how this idea was applied in the past or
how it could evolve in the future. The expansion should not simply rephrase
but create a new temporal or disciplinary perspective.

Original content: \{content\}

The text should be concise, natural, and under 150 words.

\end{tcolorbox}

During synthesis, one prompt is randomly sampled for each instance from a larger prompt pool with similar structures, encouraging diverse discourse styles and reasoning patterns while maintaining consistent length and topical alignment across the corpus.

\subsection{Details of Hard Negative Construction Prompt Design}

To enhance the discriminative capability of the dense retrieval system, we design a specialized prompt to synthesize high-quality hard negatives. Unlike simple random negatives, these instances are engineered to be topically consistent with the query but factually or logically inconsistent with the ground-truth context. The prompt instructs the MLLM to manipulate key entities, dates, or causal relationships, thereby creating plausible but incorrect candidates that challenge the model to capture fine-grained semantic differences. The specific prompt template used for hard negative generation is presented below.

\begin{tcolorbox}[
  breakable,
  enhanced,
  colback=gray!2,
  colframe=black!70,
  boxrule=0.5pt,
  arc=2pt,
  left=10pt, right=10pt, top=12pt, bottom=10pt,
  attach boxed title to top left={yshift=-3mm, xshift=5mm},
  boxed title style={
    sharp corners, 
    size=small, 
    colback=black!70, 
    colframe=black!70
  },
  title=\textbf{Hard Negative Construction Prompt},
  coltitle=white,
  fonttitle=\footnotesize\sffamily,
  fontupper=\small\ttfamily\linespread{1.1}\selectfont,
]

\textbf{[System Message]} \\
You are an expert in constructing hard negative passages for training dense retrieval systems. Your outputs must stay on the same general topic as the query, but differ in key facts or focus so they are not actually correct answers.

\textbf{[User Instruction]} \\
You are given a search query and its relevant positive passage. Write ONE hard negative passage that could plausibly be retrieved for this query but is not truly relevant or is factually inconsistent with the positive passage.

Requirements:
1. Stay on the same general topic as the query.
2. Change key facts, entities, dates, numbers, or conclusions so the passage would mislead a retrieval model if treated as a correct answer.
3. Do NOT copy or closely paraphrase sentences from the positive passage.
4. Make the passage natural, coherent, and roughly similar in length to the positive passage.
5. Do NOT explicitly mention that this is a negative or hard negative example.
6. Do NOT mention the query or the positive passage in the output.
7. Do NOT use bullet points or headings; write a single fluent paragraph.

Query: \{query\} \\
Positive: \{positive\} \\
Hard Negative:

\end{tcolorbox}

By synthesizing negatives that mirror the linguistic style and topical scope of the positive passages while introducing subtle factual conflicts, we force the retriever to move beyond shallow keyword matching and instead rely on robust semantic understanding during the fine-tuning process.

\subsection{Details of Data Statistics and Format}
\label{appendix:data_stats}

In this section, we provide a detailed breakdown of the ATIR dataset statistics and present a representative data instance to illustrate the multimodal structure used in our experiments.

\paragraph{Dataset Statistics}
The dataset is constructed by filtering and augmenting the raw corpus as described in the main paper. We finally annotate 3,909 query-positive document pairs as the test set, while the remaining 84,374 pairs constitute the training set. 

\paragraph{Data Instance Structure}
Each data instance represents a complete retrieval unit consisting of a ground-truth context, conversational Q\&A turns with aligned speech waveforms (sampled at 16kHz), and a synthesized hard negative passage. 

Below is a visualization of a representative sample from the dataset. It demonstrates how the query is grounded in the positive context ("Cotton the kitten") and contrasted with a topically distinct hard negative ("Vatican Library").


\begin{simplebox}[Representative Data Instance]
\textbf{Conversation ID:} 0

\tcblower

\textbf{\sffamily\footnotesize [Positive Context]}\par
Once upon a time, in a barn near a farm house, there lived a little white kitten named Cotton. Cotton lived high up in a nice warm place above the barn where all of the farmer's horses slept. But Cotton wasn't alone in her little home above the barn, oh no. She shared her hay bed with her mommy and 5 other sisters... \textit{(truncated for brevity)} ...Then Cotton thought, ``I change my mind. I like being special''.

\textbf{\sffamily\footnotesize [Conversational Turns]}\par
\begin{itemize}[leftmargin=*, label={--}, itemsep=2pt, topsep=2pt]
  \item \textbf{Turn 1:}
    \begin{itemize}[leftmargin=1.2em, label={}, itemsep=1pt, topsep=1pt]
      \item \textit{Question:} ``What color was Cotton?''
      \item \textit{Answer:} ``white''
      \item \textit{Audio Meta:} 16\,kHz, 0.40\,s
    \end{itemize}

  \item \textbf{Turn 2:}
    \begin{itemize}[leftmargin=1.2em, label={}, itemsep=1pt, topsep=1pt]
      \item \textit{Question:} ``Where did she live?''
      \item \textit{Answer:} ``in a barn''
      \item \textit{Audio Meta:} 16\,kHz, 0.92\,s
    \end{itemize}
\end{itemize}

\textbf{\sffamily\footnotesize [Hard Negative Passage 1]}\par
The Vatican Apostolic Library, officially known as the Vat, was established in 1523 during the reign of Pope Clement VII and serves primarily as a repository for medieval religious manuscripts and liturgical texts. Located in the heart of Vatican City, it houses approximately 900,000 printed volumes and over 60,000 manuscripts... \textit{(Note: This passage is semantically distant from the query but structurally similar to a valid document.)}

\end{simplebox}

\section{Experimental Implementation Details}
\subsection{Details of Implementation}
\label{appendix:implementation_details}

In this section, we describe the training configurations and implementation details of the ATIR framework.

\paragraph{Model and Architecture}
We adopt Qwen2.5-Omni-3B as the backbone MLLM. It is a native multimodal transformer that processes text and audio tokens within a unified architecture. The 3B-parameter scale provides a favorable trade-off between representation capacity and computational efficiency for large-scale retrieval.

\paragraph{Parameter-Efficient Fine-tuning}
To reduce memory consumption and improve training efficiency, we employ LoRA-based parameter-efficient fine-tuning~\citep{DBLP:conf/iclr/HuSWALWWC22}. LoRA adapters are inserted into the projection layers of the transformer, including the query, key, value, output, and feed-forward projections, while excluding visual-specific modules. We set the LoRA rank $r=32$, scaling factor $\alpha=32$, and dropout rate to 0.1, with Gaussian initialization and no bias parameters. During training, only LoRA parameters are updated, while the backbone model remains frozen. For Stage~II training, we initialize the model from the LoRA checkpoint obtained in Stage~I and continue fine-tuning the same adapters.

\paragraph{Training Hyper-parameters}
Both training stages are run for \textbf{2 epochs}. We use the AdamW optimizer~\citep{DBLP:conf/iclr/LoshchilovH19} with a peak learning rate of $5\times10^{-5}$ and a linear warm-up schedule over the first $10\%$ of training steps (capped at 500 steps). The temperature parameter $\tau$ in the contrastive loss is set to 0.05. Gradient accumulation and gradient checkpointing are applied to support large effective batch sizes under limited GPU memory.

\paragraph{Hardware and Software}
All experiments are conducted on a cluster of 8$\times$ NVIDIA A100 GPUs (40GB). We use DeepSpeed with ZeRO optimization for distributed training and memory efficiency. The maximum text sequence length is set to 512 tokens, while audio inputs are processed using the native Qwen2.5-Omni tokenizer with a sampling rate of 16kHz. The full two-stage training process takes approximately 24 hours.

\subsection{Details of Baseline}

For Omni-Embed-Nemotron-3B, we reproduce the model using the official codebase and training configuration provided by the authors.\footnotemark[3]\footnotetext[3]{\url{https://huggingface.co/nvidia/omni-embed-nemotron-3b}} Following the original setup, the reproduced model achieves expected performance on single-modality retrieval tasks. However, when evaluated under cross-modal, fused-modal, and audio--text interleaved retrieval settings, its performance degrades significantly, indicating limited generalization to more complex multimodal retrieval scenarios.

To better adapt the model to ATIR task, we further fine-tune Omni-Embed-Nemotron-3B on our dataset using the official LoRA-based training strategy. Specifically, we freeze the audio and visual encoders and apply LoRA tuning only to the language model, as suggested in the original work. We adopt the recommended hyperparameters (LoRA rank $r=16$ and scaling factor $\alpha=32$) and retain the bidirectional attention modification. This additional training stage allows the model to better align audio and text representations under our retrieval setting, yielding improved but still inferior performance compared to ATIR-Qwen-3B.

\section{Usage of AI Assistants}
We use ChatGPT to improve the presentations of this paper.\footnotetext[4]{\url{https://chatgpt.com/}}

\end{document}